\pdfoutput=1

\documentclass[11pt]{article}

\usepackage[]{ACL2023}

\usepackage{times}
\usepackage{latexsym}

 \usepackage{graphicx} 
 \usepackage{float}
 \usepackage{booktabs}

\usepackage[T1]{fontenc}

\usepackage[utf8]{inputenc}

\usepackage{microtype}

\usepackage{inconsolata}

\title{Open Political Corpora: Structuring, Searching, and Analyzing Political Text Collections with PoliCorp}

\author{Nina Smirnova \and Muhammad Ahsan Shahid \and Philipp Mayr\\
          GESIS – Leibniz Institute for the Social Sciences \\ Unter Sachsenhausen 6-8, 50667 Cologne \\
 \texttt{\{nina.smirnova, ahsan.shahid, philipp.mayr\}@gesis.org}
 } 

\begin{document}
\maketitle
\begin{abstract}
In this work, we present \href{https://demo-pollux.gesis.org/}{PoliCorp}, a web portal designed to facilitate the search and analysis of political text corpora. PoliCorp provides researchers with access to rich textual data, enabling in-depth analysis of parliamentary discourse over time.  The platform currently features a collection of transcripts from debates in the German parliament, spanning 76 years of proceedings. With the advanced search functionality, researchers can apply logical operations to combine or exclude search criteria, making it easier to filter through vast amounts of parliamentary debate data. The search can be customised by combining multiple fields and applying logical operators to uncover complex patterns and insights within the data. Additional data processing steps were performed to enable web-based search and incorporate extra features. A key feature that differentiates PoliCorp is its intuitive web-based interface that enables users to query processed political texts without requiring programming skills. The user-friendly platform allows for the creation of custom subcorpora via search parameters, which can be freely downloaded in JSON format for further analysis.
\end{abstract}

\section{Introduction}

Parliamentary debates offer a broad variety of topics for academic exploration, serving as an effective instrument for agenda-setting and influencing political power. By examining parliamentary speeches, researchers can uncover the implicit programmatic and ideological positions political parties hold. Recent studies include investigations into gender dynamics and examinations of negativity levels inferred from interjections \citep{ash_gender_2024} and sentiment and negativity analysis \citep{Jenny_Marcelo_2021}. Additional efforts focus on developing automated topic modeling \citep{doi:10.1177/0894439320907027} and discursive framing approaches within legislative settings \citep{reinig-etal-2024-politics}.


Parliamentary data from many countries is openly available online through governmental initiatives and open data platforms. Examples include the Open Data portal of the German parliament (Bundestag)\footnote{\url{https://www.bundestag.de/services/opendata}}, the Austrian parliament transcripts\footnote{\url{https://www.data.gv.at/en/}}, and Hansard, the official record of the UK Parliament\footnote {\url{https://hansard.parliament.uk/}}. In addition to raw data, some initiatives provide preprocessed and linguistically annotated corpora, e.g.,  the Polish Parliamentary Corpus\footnote{\url{https://clip.ipipan.waw.pl/PPC}}, GermaParl, a linguistically annotated corpus of German Bundestag plenary debates \citep{germaparl_r, andreas_blaette_germaparl_2017}, and DutchParl, which contains documents from the parliaments of the Netherlands, Flanders, and Belgium in Dutch \citep{marx-schuth-2010-dutchparl_v2}. Several platforms offer web-based search interfaces for parliamentary records, such as the polit-X portal\footnote{\url{https://polit-x.de/}}, which includes transcripts from the German Bundestag and state parliaments; the StateParl portal\footnote{\url{https://stateparl.de/}}, which covers debates from the 16 German state parliaments; and the Italian parliamentary transcripts portal\footnote{\url{https://accademiadellacrusca.it/it/contenuti/discorsi-parlamentari/23591}}. 

Although these resources provide comprehensive collections, they often present certain limitations. Raw data typically requires significant preprocessing before it can be analyzed, and many annotated datasets are distributed as complete corpora, necessitating advanced analytical skills to extract specific information \citep{marx-schuth-2010-dutchparl_v2, andreas_blaette_germaparl_2017}. Furthermore, some data sets are embedded within specialized software environments, which require a knowledge of particular programming languages for effective use \citep{germaparl_r, blatte_polminer_2020}. Although certain platforms, such as polit-X, offer user-friendly online interfaces for data exploration, they often operate on a subscription basis, limiting open access.
Overall, the lack of standardized, machine-readable annotations for parliamentary speeches limits the potential for quantitative research in this area \citep{wissik_encoding_2021}.

To overcome these challenges, we introduce the Pollux Political Corpora (PoliCorp) platform\footnote{\url{https://demo-pollux.gesis.org/}}, an advanced resource that offers researchers open, structured, and searchable access to processed political corpora. The platform currently contains a collection of transcripts of Bundestag debates, spanning 76 years of parliamentary debates -- from September 1949 to July 2025.  
The platform is designed for political scientists, interdisciplinary researchers, and others engaged in the analysis of parliamentary discourse.

Additional data processing steps were performed to enable web-based search and incorporate supplementary features.
For instance, PoliCorp allows users to perform targeted searches not only within speeches but also across "interjections" and "calls to order". 
Calls to order can serve as indicators for analyzing incivility in parliamentary discourse, and offer a unique perspective on political polarization \citep{Jenny_Marcelo_2021}, and are therefore of particular interest for political research. Moreover, analysis of calls to order as markers of disruptive language is a novel approach to studies of parliamentary corpora, going beyond traditional sentiment or stance analysis. 
Interjections constitute a key resource for understanding democratic processes, uncovering hidden power dynamics, and examining conflicts within parliament \citep{ilie_parliamentary_2015, truan_pragmatics_2017}.

A key feature that differentiates PoliCorp is its intuitive web-based interface that enables users to query processed political texts without requiring programming skills\footnote{A demonstration of the search functionality is available at the following link: \url{https://youtu.be/KplgIZRVVwQ}}. The user-friendly platform allows the creation of custom subcorpora via search parameters, which can be freely downloaded in JSON format for further analysis. The portal’s content is distributed under the CLARIN PUB+BY+NC+SA license.



\subsection*{Background on Parliamentary Discourse}

The legislative period is a specific period of time for which a parliament is elected. In Germany, it spans four years.
Within each legislative period, the work of the German Bundestag is organized through plenary sessions, which follow a set procedure.  
The session is presided over by a member of the Bundestag Presidium, comprising the president and two secretaries. The president moderates the session, ensuring procedural compliance and managing the allocation of speaking time. Speaking time is distributed proportionally among parliamentary groups, reflecting their relative representation in the parliament. 
Each session usually comprises several agendas, dedicated to a specific topic.

If the speaker or an attending parliamentary member violates parliamentary order, i.e., interruptions, the president is authorized to issue a formal call to order\footnote{\url{https://www.bundestag.de/services/glossar/glossar/O/ordnungsruf-869614}}. Calls to order were analysed by \cite{Jenny_Marcelo_2021} from a perspective of negativity analysis. Figure~\ref{fig:cto_policrop} demonstrates an example of a call to order during a parliamentary session from the PoliCorp interface.

Interruptions or interjections during a speech may include spontaneous remarks or attempts to insert commentary by other members of parliament\footnote{\url{https://de.wiktionary.org/wiki/Zwischenruf}}.  Interjections are widely examined in political science research. 
Research on interjections has largely addressed their pragmatic, rhetorical, and disruptive functions \citep{truan_pragmatics_2017, shenhav_showing_2008}.   
Many scholarly works have further investigated interjections in relation to gendered dynamics of interjections, including gendered patterns of attacks, interruptions, and participation in legislative debates across different contexts \citep{van_dijk_interrupting_2025, ash_gender_2024, vallejo_vera_politics_2022, och_manterrupting_2020, poljak_role_2022, vallejo_vera_politics_2022, MILLER_SUTHERLAND_2023}. 
Additional lines of inquiry have explored interjections in relation to expertise, seniority, and affiliation dynamics \citep{diener_explaining_2025}, cross-cultural variation in speech styles \citep{isosavi_reactions_2025}, and broader patterns of interruption behavior \citep{poljak_parties_2023}.

\section{Backend Implementation}

Raw parliamentary speeches up to September 7, 2021, were sourced from the GermaParl corpus \citep{andreas_blaette_germaparl_2017}, a comprehensive linguistic dataset curated by the PolMine project\footnote{\url{https://polmine.github.io/}}. GermaParl covers transcripts of parliamentary debates from September 7, 1949, to September 7, 2021, and comprises of 958,100 speech contributions. Raw parliamentary speeches published after September 7, 2021, were sourced from the Bundestag Open Data project\footnote{\url{https://www.bundestag.de/services/opendata}}.
The raw GermaParl data is available for download in a  GitHub repository\footnote{\url{https://github.com/PolMine/GermaParlTEI/tree/main}}. The records from the Bundestag Open Data project were retrieved using an API. 

\begin{figure}[h!]
\centering
  \includegraphics[width=0.8\columnwidth]{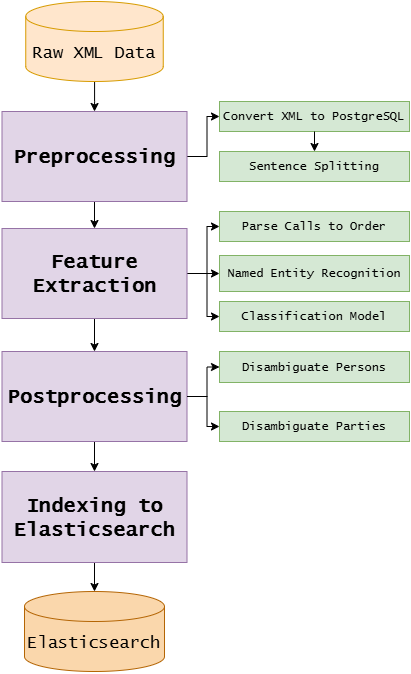}
  \caption{Data processing workflow for PoliCorp}
  \label{fig:preprocessing_worklow}
\end{figure}

The raw data underwent a series of processing stages, as illustrated in Figure~\ref{fig:preprocessing_worklow}. The full processed dataset, currently hosted by PoliCorp, comprises 1,035,744 speech contributions\footnote{As of September 2025.}. In the final stage, the processed data is indexed using Elasticsearch (version 8.12.1), facilitating efficient retrieval of user-specific information via web-interface.  

\subsection{Preprocessing}\label{sec:preprocessing}

The Bundestag and GermaParl data were originally available in XML format, although with different structural schemas (see Appendix~\ref{sec:raw_xml}). To address this issue, we developed two distinct conversion pipelines to map each XML structure into a unified database schema. The unified data schema comprises information about the agenda, speaker, and a corresponding speech contribution, as Figure~\ref{fig:metadata_demo} demonstrates. After the conversion process, the textual content of individual speech contributions was segmented into sentences.

\subsection{Feature Extraction}\label{sec:feature_extraction}

\begin{figure}[h]
\centering
  \includegraphics[width=0.8\columnwidth]{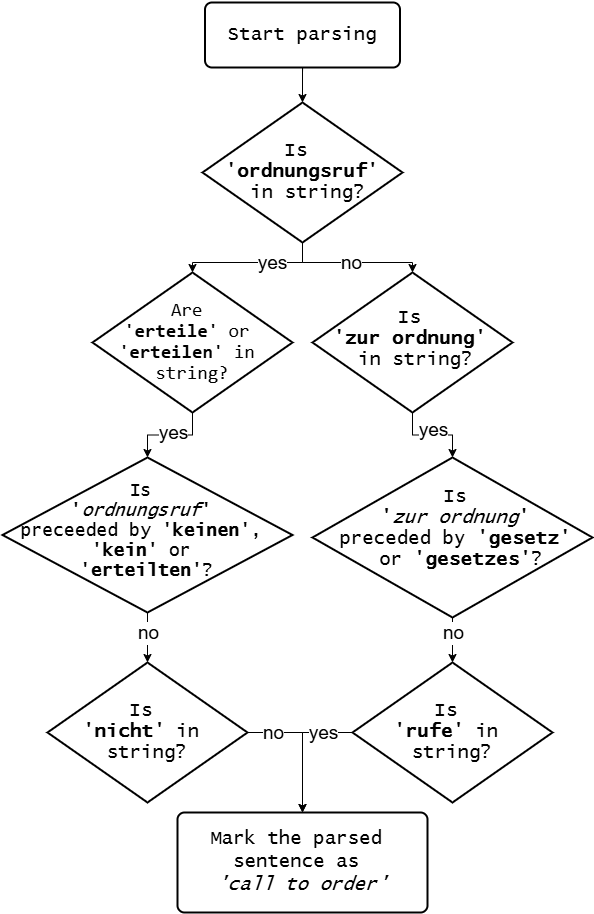}
  \caption{Rules to match calls to order}
  \label{fig:cto_rules}
\end{figure}

\begin{figure*}[h!]
\centering
 \includegraphics[width=1\linewidth]{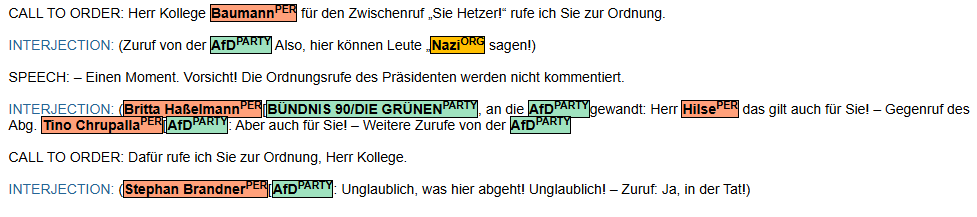}
  \caption{Example of marked call to order and NER output from the PoliCorp interface in the speech of the session's president, Wolfgang Schäuble, on October 2, 2020.}
  \label{fig:cto_policrop}
\end{figure*}

Bundestag specifically marks interjections in their raw data (as Figures~\ref{fig:raw_germaparl} and \ref{fig:raw_bundestag} demonstrate). Conversely, calls to order are not explicitly indicated in the data. Our objective was also to identify sentences that include calls to order. This procedure is regulated and involves the use of specific phrases. In the first step, we manually reviewed a part of the dataset containing only the speeches of the session's president. Based on this review, we developed a set of rules to identify calls to order, as illustrated in Figure~\ref{fig:cto_rules}. Following, we applied these rules to analyze only the speeches given by the session's president to detect instances of calls to order. Figure~\ref{fig:cto_policrop} shows an example of a marked call to order in the PoliCorp interface\footnote{Translation of the text in Figure~\ref{fig:cto_policrop} is provided in Appendix~\ref{sec:trasnlation}}.

The output of the Named Entity Recognition (NER)  models is integrated into PoliCorp as an experimental feature.  Currently, users can see the output of two NER models, Legal German NER\footnote{\url{https://huggingface.co/flair/ner-german-legal}} \citep{leitner2019fine, akbik2018coling} and German NER\footnote{\url{https://huggingface.co/flair/ner-german}} \citep{akbik2018coling}. We conducted additional processing of the models' output and retrieved mentions of the German parties using pattern matching. Figure~\ref{fig:cto_policrop} shows the example of the output of the German NER model in the PoliCorp interface.

\begin{figure}[h!]
\centering
 \includegraphics[width=\columnwidth]{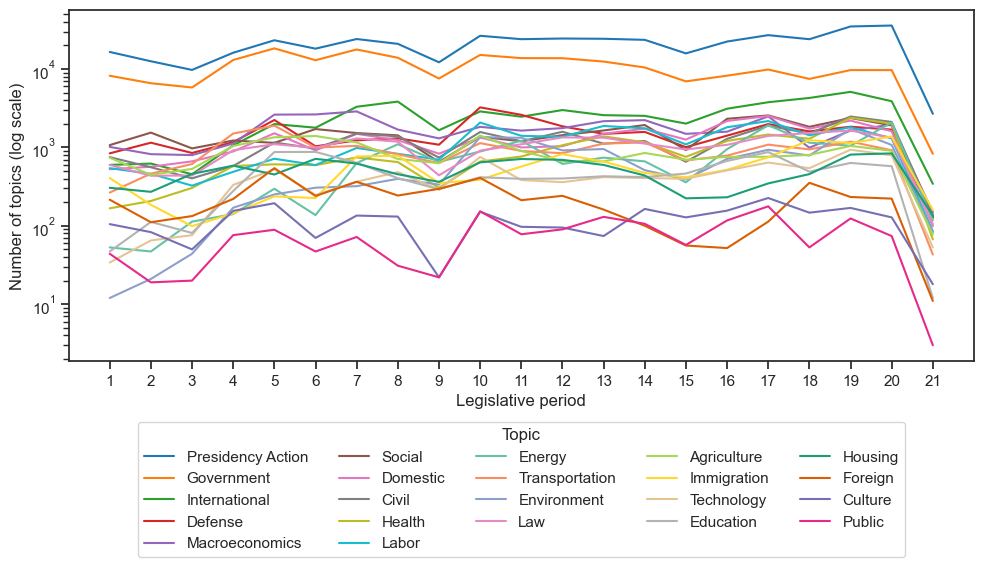}
  \caption{Distribution of topics over legislative periods.}
  \label{fig:topics}
\end{figure}

In addition, each speech contribution is annotated with a topic classification, generated using a BERT-based model\footnote{\url{https://huggingface.co/chkla/parlbert-topic-german}} \citep{klamm2022frameast}. The classifier was applied to the full text of each speech contribution, excluding those delivered by the session president, which merely consist of procedural moderation. The underlying model was trained to differentiate between 21 distinct topics relevant to German parliamentary discourse. Contributions from the session president were instead assigned a separate category, labeled Presidency Action, thereby introducing an additional topic class. Figure~\ref{fig:topics} shows the distribution of topics in the PoliCorp collection over 21 legislative periods. The diagram illustrates the longitudinal development of the discussed topics within the Bundestag. Across all legislative periods, presidency actions and governmental affairs emerge as the most consistently debated and stable topics. In contrast, the environment topic, while being of comparatively low relevance during the early legislative periods, progressively gained relevance, becoming one of the major topics in the late periods. 

\subsection{Postprocessing}\label{sec:postprocessing} 

As the final processing step, the names of speakers and their associated political parties were disambiguated. This step was necessary due to inconsistencies in the raw data, which included spelling errors in speaker names, variations of the first names, as well as multiple variations of political party names, i.e., both abbreviated and full forms. 

For the disambiguation of speaker names, a rule-based approach was employed. This approach utilized a database containing the names of all members of the German Parliament throughout its history\footnote{\url{https://www.bundestag.de/services/opendata}}. 
Individuals with unique surname or surname–first name combinations were directly assigned the corresponding identifier. In cases where identical surname or surname–first name combinations appeared multiple times in the database, additional disambiguation was necessary. This was addressed by aligning the individuals with the legislative periods during which the corresponding speech contribution was delivered. If the date of a speech fell within an individual's documented parliamentary tenure, that person was considered a match. When multiple potential matches remained or no corresponding surname–first name combination could be identified in the database, the entry was kept as ambiguous, and no identifier was assigned.

Subsequently, political party names were disambiguated using pattern matching in conjunction with a comprehensive list of all German political parties and their known abbreviations across the history of the Bundestag. This process allowed for the normalization of party references into a unified format. For instance, both Freie Demokratische Partei and FDP were identified as referring to the same political entity. 

\section{Frontend Implementation}

The frontend is developed using Express.js with Pug templates for server-side rendering. It provides an interactive web-based interface for querying, visualizing, and downloading results from parliamentary corpora. Users can perform both simple and advanced queries, with input parameters dynamically translated into Elasticsearch queries targeting the backend index. The interface includes additional informational pages (e.g., About, GitHub, FAQs, tutorial videos) to support usability. The design emphasizes responsiveness, modularity, and seamless backend integration to enable efficient and effective information retrieval for political science research.

\subsection{User Interface}
\begin{figure}[h]
\centering
  \includegraphics[width=\columnwidth]{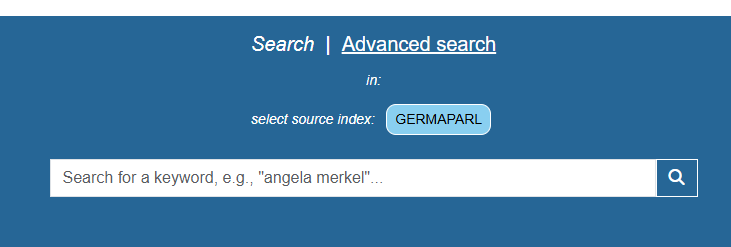}
  \caption{Basic search}
  \label{fig:demo_search}
\end{figure}

\begin{figure}[h]
\centering
  \includegraphics[width=\columnwidth]{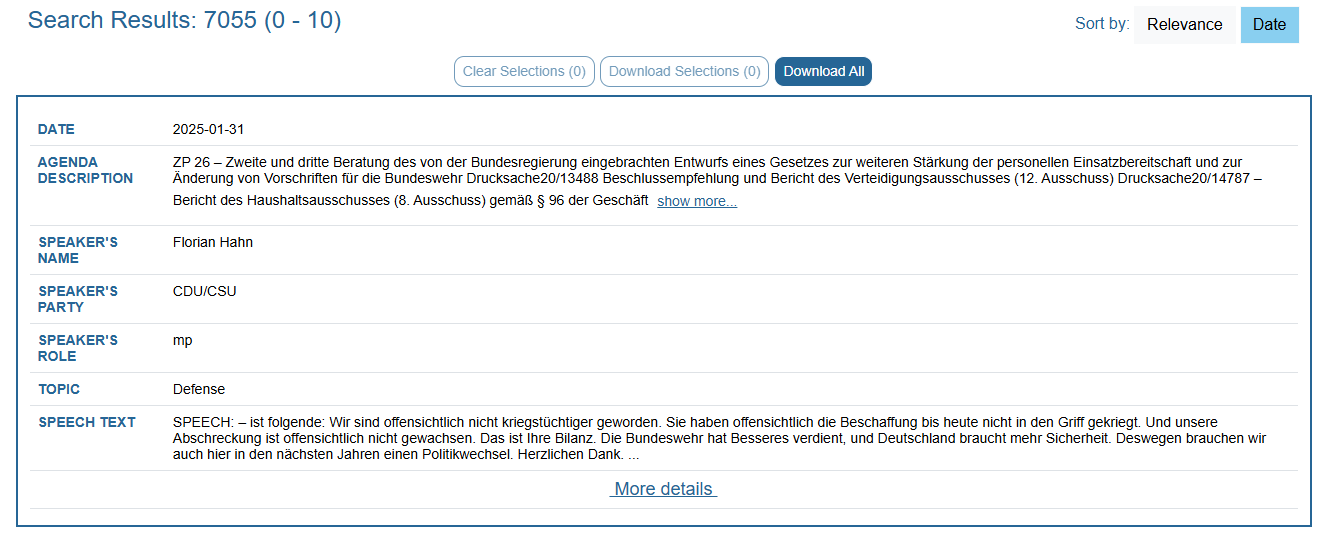}
  \caption{Search results}
  \label{fig:search_res}
\end{figure}

Figure~\ref{fig:demo_search} represents a basic search bar. By default, the query is executed against the full text of a speech contribution and all indexed metadata fields. Upon submission, the system returns a results list, displaying the total number of matches and an expandable metadata panel, as Figure~\ref{fig:search_res} shows. Results may be reordered by relevance or chronological order.

\begin{figure}[h]
\centering
  \includegraphics[width=\columnwidth]{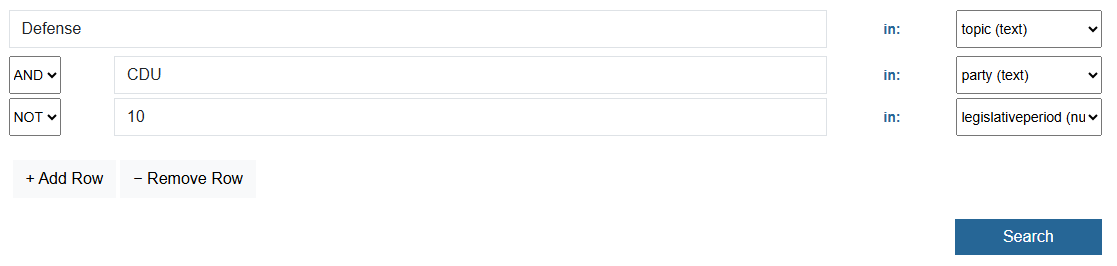}
  \caption{Advanced search}
  \label{fig:advanced_search}
\end{figure}

For a more specific search, the user can proceed by selecting an advanced search option, as Figure~\ref{fig:advanced_search} shows. The advanced interface has a set of structured input rows, each row comprising three elements: a logical operator, a field selector, and a value field.
Logical operator fields may be configured individually as AND, OR, or NOT, permitting execution of complex Boolean expressions. The drop-down list with the field selectors contains all searchable attributes, including but not limited to: full text, speaker, party, legislative period, topic, or date.

\begin{figure}[h]
\centering
  \includegraphics[width=\columnwidth]{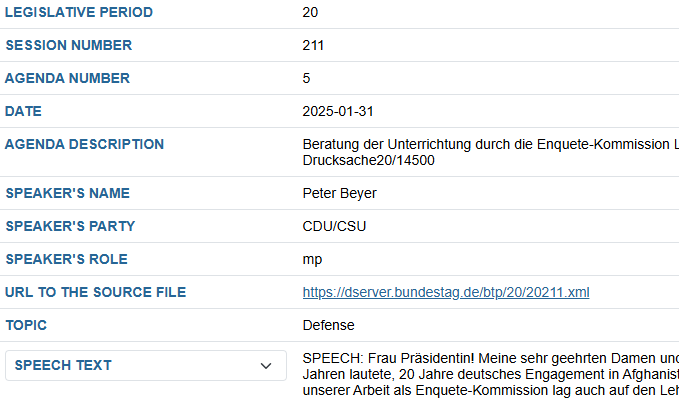}
  \caption{Available metadata}
  \label{fig:metadata_demo}
\end{figure}

If a user is interested in a specific speech contribution, an expandable metadata panel can be accessed. Figure~\ref{fig:metadata_demo} displays the complete set of metadata associated with the selected speech, including general session-related information such as the legislative period, agenda and session sequence numbers, the date of the session, as well as the agenda type and a brief description. Additional metadata covers speaker-related details, including the speaker’s name, party affiliation, and role or position within the Bundestag. Furthermore, the topic of the speech and a link to the corresponding source file are also provided. The web portal incorporates experimental features, such as the outputs of two NER models, as detailed in  Section~\ref{sec:feature_extraction}. These features are accessible via a drop-down menu labeled Speech Text. Selecting a specific NER model from the list dynamically updates the display of the speech contribution, as illustrated in Figure~\ref{fig:cto_policrop},  by highlighting the recognized named entities within the text.  

\begin{figure}[h]
\centering
  \includegraphics[width=\columnwidth]{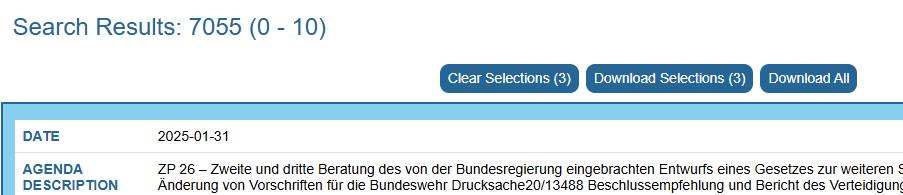}
  \caption{Download options}
  \label{fig:download_demo}
\end{figure}

A key feature of PoliCorp is that it enables users to combine data in various ways, i.e., generating custom subcorpora and exporting them in JSON format (Figure~\ref{fig:download_demo}). 

\begin{figure}[h]
\centering
  \includegraphics[width=\columnwidth]{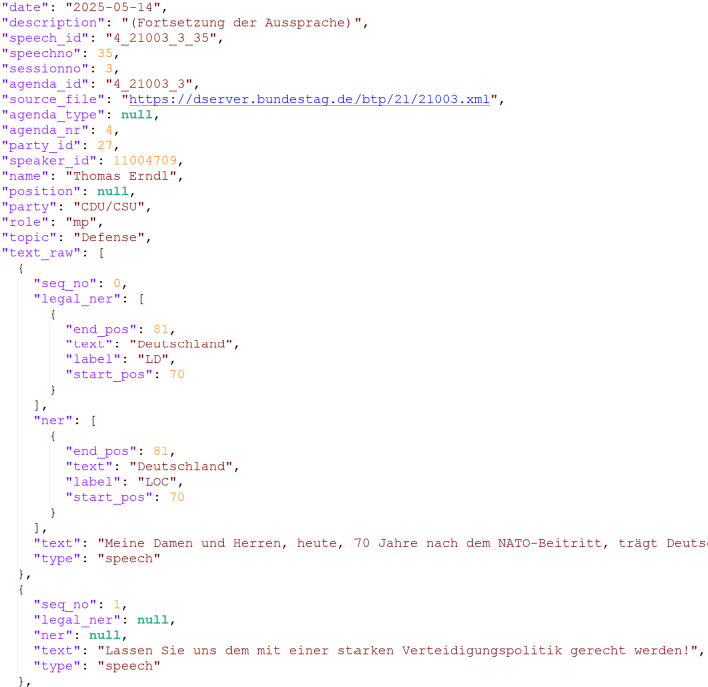}
  \caption{Example of JSON format }
  \label{fig:download_json}
\end{figure}

The data contained in the downloaded JSON file, as shown in Figure~\ref{fig:download_json}, fully corresponds to the information presented in the web interface. It includes all associated metadata, sentence-level segmentation of the speech contribution, and available named entity annotations.
Thus, metadata comprises details about the speaker, including their identity, affiliation, and functional role. Furthermore, it encompasses information specific to the speech itself, such as the legislative period, session, and agenda numbers, date of delivery, and the speech topic.

\section{Portal Usage and Evaluation}

Between January 5, 2025, and September 14, 2025, a total of 2,573 user queries were executed within the PoliCorp system, comprising 699 unique queries. The statistical analysis excludes queries exhibiting patterns indicative of potential malicious activity, such as those containing keywords like system, sleep, exec, bash, and similar commands. Table~\ref{tab:stat_usage} displays the top 10 most frequently searched terms. 

\begin{table}[h]
\centering
\begin{tabular}{lr}
\toprule
search term & frequency \\
\midrule
19 & 118 \\
20 & 93 \\
cdu & 89 \\
merkel & 77 \\
spd & 76 \\
angela merkel & 66 \\
atomausstieg & 55 \\
klimawandel & 52 \\
migration & 39 \\
die linke & 35 \\
\bottomrule
\end{tabular} 
\caption{Top 10 Most Frequently Searched Terms}
\label{tab:stat_usage}
\end{table}

Statistics indicate that users typically search for data from a specific legislative period, e.g, 19 and 20. The second most popular search requests include political parties (CDU, SPD, DIE LINKE) and names of politicians (Merkel, Angela Merkel). The other common search requests relate to keywords such as Atomausstieg (nuclear phase-out), Klimawandel (climate change), or  Migration (migration). 

The evaluation of the web portal focused on user-centered design principles. Functionality of PoliCorp was developed and refined in collaboration with a small user group with a background in political science or computer science-related disciplines, ensuring that key features align with user needs and expectations. Based on the evaluation results, certain interface and analysis features were added or removed from the portal, and the final interface layout was developed.

\section{ Conclusion and Future Work}

In this work, we present PoliCorp, a web portal designed to facilitate the search and analysis of political text corpora. PoliCorp provides researchers with access to rich textual data, enabling in-depth analysis of parliamentary discourse over time.  The platform currently contains a collection of transcripts of Bundestag debates, spanning 76 years of parliamentary debates. With the advanced search functionality, researchers can apply logical operations to combine or exclude search criteria, making it easier to filter through vast amounts of parliamentary debate data. The search can be customised by combining multiple fields and applying logical operators to uncover complex patterns and insights within the data. Selected datasets can be downloaded freely in JSON format, providing a convenient option for further analysis using computational tools.

PoliCorp is a demonstration version that is currently under development. 
As of the current period, the platform comprises speech contributions from the German Bundestag covering the period from September 1949 to July 2025. Its content will be continuously updated with new speeches as they become available through the Bundestag Open Data portal. Ongoing work includes the implementation of a toxicity detection module, which will assign predefined toxicity labels to individual speech segments. Following, we are planning to link available transcripts with corresponding video recordings. 

Future iterations will incorporate additional corpora, such as StateParl, and present them in a unified format to facilitate consistent processing and comparative analysis.

Furthermore, we are working on the integration of the analysis tools, i.e., descriptive statistics and visualization, and cross-corpus content comparison capabilities. Additionally, the available download formats will be expanded. The tool’s source code will be made publicly available in a future release.

\section*{Limitations}

PoliCorp represents a prototype implementation, with many features still under active development. Due to the early stage of the project, the web portal has been evaluated from a user-centered design perspective using a small group of participants. Broader evaluations, including surveys and workshops, are planned for future phases.

The platform incorporates experimental functionalities such as topic classification and NER, both of which are generated automatically and may contain inaccuracies. Users are therefore advised to interpret these outputs with caution and independently verify any critical information.

Furthermore, PoliCorp is a domain-specific resource primarily intended for users engaged in political science research.


\section*{Acknowledgements}

Nina Smirnova received funding from the Deutsche Forschungsgemeinschaft (DFG) under grant number: MA 3964/7-3 (POLLUX Project).

Nina Smirnova and Philipp Mayr received additional funding from the European Union under the Horizon Europe grant OMINO \cite{omino2024} – Overcoming Multilevel Information Overload (\url{http://ominoproject.eu}) under grant number 101086321.

Muhammad Ahsan Shahid received funding from the Deutsche Forschungsgemeinschaft (DFG) under grant number: MA 3964/20-1 (OFFZIB Project).

\bibliography{anthology,custom}
\bibliographystyle{acl_natbib}

\appendix

\onecolumn

\section{Raw XML Formats}\label{sec:raw_xml}

\begin{figure}[H]
\centering
  \includegraphics[width=\textwidth]{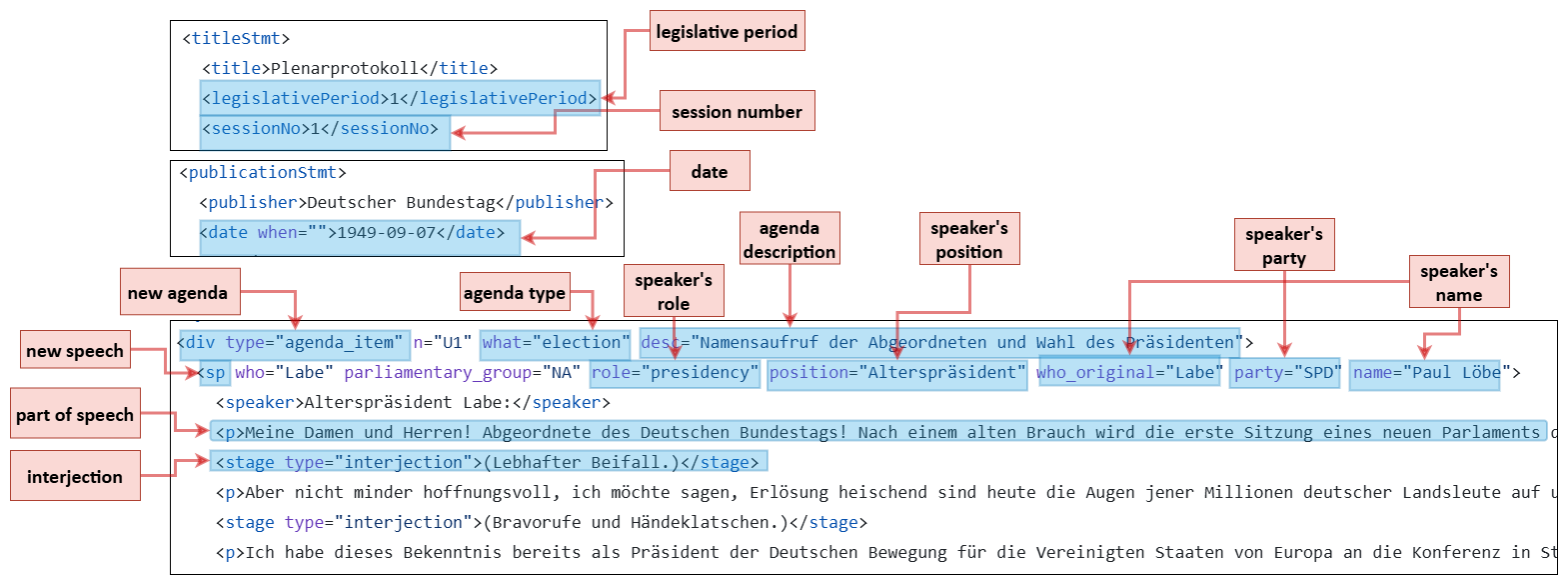}
  \caption{Example of the raw GermaParl data format}
  \label{fig:raw_germaparl}
\end{figure}

\begin{figure}[H]
\centering
  \includegraphics[width=\textwidth]{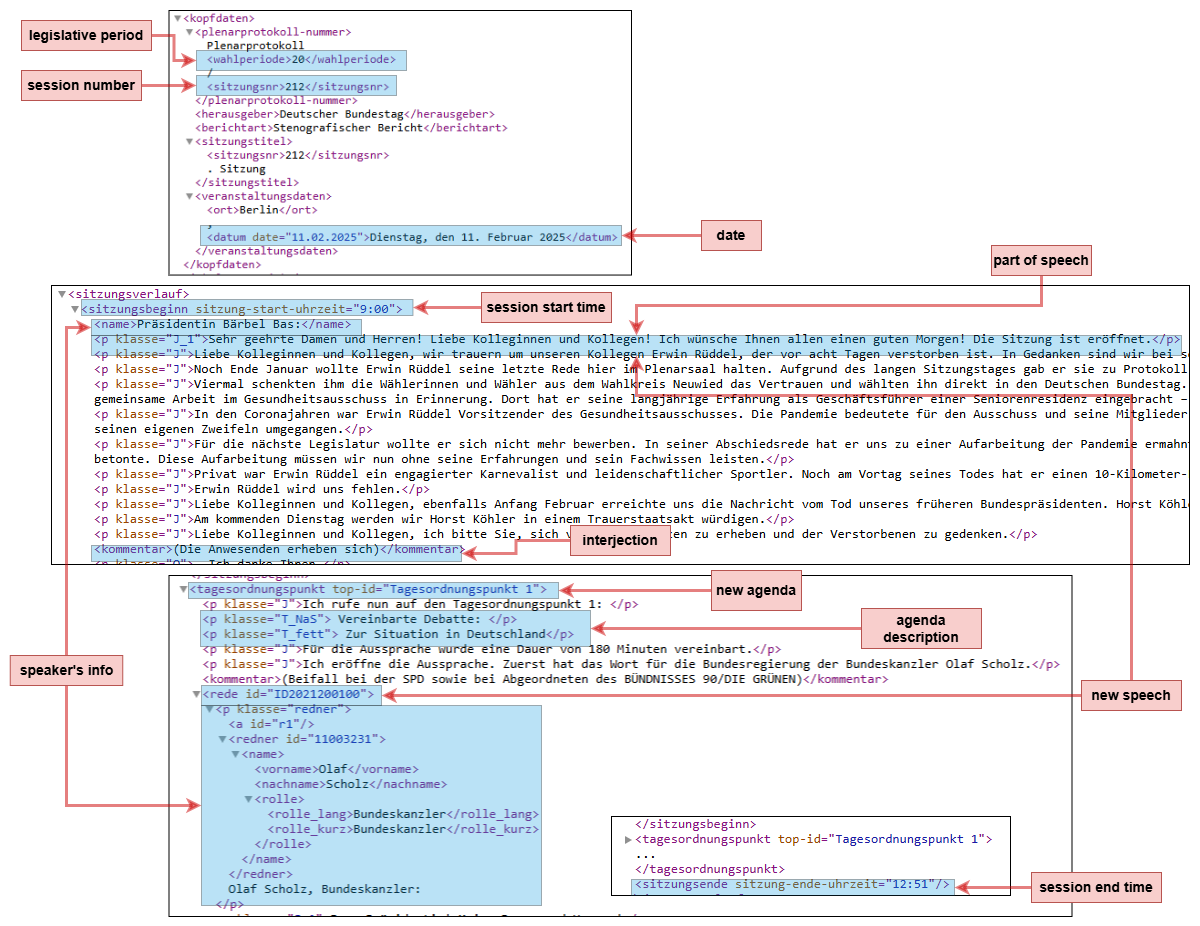}
  \caption{Example of the raw Bundestag Open Data data format}
  \label{fig:raw_bundestag}
\end{figure}

\section{Translation of the German text in Figure~\ref{fig:cto_policrop}}\label{sec:trasnlation}

\begin{itemize}
    \item \textbf{CALL TO ORDER:} Mr. Baumann, I call you to order for the interjection “You agitator!”.
    \item \textbf{INTERJECTION:} (Shout from the AfD: So, people can say “Nazi” here!)
    SPEECH: - One moment. Watch out! The President's calls to order will not be commented on.
    \item \textbf{INTERJECTION:} (Britta Haßelmann [Alliance 90/The Greens], addressing the AfD: Mr. Hilse, that goes for you too! - Counter-cry from Tino Chrupalla [AfD]: But also for you! - More shouts from the AfD)
    \item \textbf{CALL TO ORDER:} I call you to order for that, Mr. Hilse.
    \item \textbf{INTERJECTION:} (Stephan Brandner [AfD]: Unbelievable what's going on here! Unbelievable! - Shout-out: Yes, indeed!)
    
\end{itemize}

\end{document}